\documentclass[11pt]{article}
\usepackage{amssymb}
\usepackage{epsfig}

\newcommand{\BE}{\begin{equation}}
\newcommand{\EE}{\end{equation}}
\begin{document}
\begin{titlepage}

\vspace*{1mm}
\begin{center}

\LARGE
   {\LARGE{\bf Gravity as an emergent phenomenon:\\ experimental signatures }}

\vspace*{10mm} {\Large  M. Consoli$^a$ and A. Pluchino$^{a,b}$ }
\vspace*{8mm}\\
{\large
a) Istituto Nazionale di Fisica Nucleare, Sezione di Catania \\
b) Dipartimento di Fisica e Astronomia dell' Universit\`a di Catania
\\}
\end{center}
\vspace*{5mm}
\begin{center}
{\bf Abstract}
\end{center}

\noindent

\par\noindent  According to some authors, gravity might be
an emergent phenomenon in a fundamentally flat space-time. In this
case the speed of light in the vacuum would not coincide exactly
with the basic parameter $c$ entering Lorentz transformations and,
for an apparatus placed on the Earth's surface, light should exhibit
a tiny fractional anisotropy at the level $10^{-15}$. We argue that,
most probably, this effect is now observed in many room-temperature
ether-drift experiments and, in particular, in a very precise
cryogenic experiment where this level of signal is about 100 times
larger than the designed short-term stability. To fully appreciate
what is going on, however, one should consider the vacuum as a true
physical medium whose fundamental quantum nature gives rise to the
irregular, non-deterministic pattern which characterizes the
observed signal.

%\vskip
%35 pt PACS: 03.30.+p, 01.55.+b
\end{titlepage}

{\bf 1.}~ The usual interpretation of gravitational phenomena is in
terms of a universal metric field $g_{\mu\nu}(x)$ viewed as a
fundamental modification of flat space-time. Reconciling this
interpretation with some basic aspects of the quantum theory may
pose, however, some consistency problems, see e.g. \cite{sonego1}
and references quoted therein. For this reason, one could try to
explore a different approach where instead curvature is an emergent
phenomenon \cite{barcelo1,barcelo2} in flat space-time analogously
to the curvature of light in Euclidean space when propagating in a
medium with variable density. This point of view may become natural
if, by taking seriously the phenomenon of vacuum condensation in
particle physics, the vacuum starts to be considered a true
superfluid medium \cite{volo}, i. e. a quantum liquid
\cite{plafluid}.

As a definite scenario, an effective metric tensor $g_{\mu\nu}(x)$
could then originate from local modifications of the basic
space-time units and of the velocity of light which are known, see
e.g. \cite{feybook,dicke,atkinson}, to represent an alternative way
to introduce the concept of curvature \footnote{This point of view
has been vividly represented by Thorne in one of his books
\cite{thorne}: "Is space-time really curved ? Isn't it conceivable
that space-time is actually flat, but clocks and rulers with which
we measure it, and which we regard as perfect, are actually rubbery
? Might not even the most perfect of clocks slow down or speed up
and the most perfect of rulers shrink or expand, as we move them
from point to point and change their orientations ? Would not such
distortions of our clocks and rulers make a truly flat space-time
appear to be curved ? Yes". }. The only possibly new aspect is that
the scale over which $g_{\mu\nu}(x)$ varies (say a small fraction of
millimeter or so) is taken much larger than any elementary particle,
or nuclear, or atomic size \cite{ultraweak}. In this sense, the type
of description of classical general relativity, and of its possible
variants, becomes similar to hydrodynamics that, concentrating on
the properties of matter at scales larger than the mean free path
for the elementary constituents, is insensitive to the details of
the underlying molecular dynamics.

By following this interpretation, one could first consider the
simplest two-parameter scheme of an effective isotropic metric \BE
\label{first} {g}_{\mu \nu}= {\rm diag}(A ,-B,-B,-B)\EE This depends
on two functions which, in a flat-space picture, can be interpreted
in terms of an overall re-scaling $\lambda$ of the space-time units
and of a refractive index ${\cal N}$ so that $A=c^2
{{\lambda^2}\over{{\cal{N}}^2}}$ and $B=\lambda^2$. Now, since
physical units of time scale as inverse frequencies, and the
measured frequencies $\hat \omega$ for a Newtonian potential $ U_N
\neq 0$ are red-shifted when compared to the corresponding value
$\omega$ for $U_N = 0$, this fixes the value of $\lambda$.
Furthermore, independently of any specific underlying mechanism, the
two functions $A$ and $B$ can be related through the general
requirement $AB=c^2=$ constant which expresses the basic property of
light of being, at the same time, a corpuscular and undulatory
phenomenon \cite{ultraweak}. This fixes the value of ${\cal N}$
giving the structure \BE \label{second} {\cal{N}}\sim
1+2{{|U_N|}\over{c^2}}~~~~~~~~~~~~~~ ~~~~~ ~~~~~~~\lambda\sim
1+{{|U_N|}\over{c^2}}\EE which to first order is equivalent to
general relativity. Finally more complicated metrics with
off-diagonal elements $g_{0i}\neq 0$ and $g_{ij}\neq 0$ can be
obtained by applying boosts and rotations to Eq.(\ref{first}). This
basically reproduces the picture of the curvature effects in a
moving fluid with a metric tensor which depends on
$\varphi={{U_N}\over{c^2}}$ in a definite parametric form, i.e. $
g_{\mu \nu}(x)=g_{\mu \nu}[\varphi (x)]$. In this way, a first
consistency check of an emergent-gravity approach consists in
constructing some long-wavelength vacuum excitation $\varphi(x)$
that, on a coarse grained scale, behaves as the Newtonian potential
\cite{ultraweak} \footnote{At a classical level, particle
trajectories in this field do not depend on the particle mass thus
allowing to establish an analogy between the motion of a body in a
gravitational field and the motion of a body not subject to an
external field but viewed by a non-inertial observer. In an
emergent-gravity approach, this is the path to new, approximate
forms of physical equivalence, i.e. different from the basic Lorentz
group. These forms are not postulated from scratch, as in general
relativity, but originate from the underlying vacuum structure.}.

Being faced with two alternative interpretations, one may wonder
whether the basic conceptual difference with standard general
relativity could have phenomenological implications \footnote{Here
we will only consider the limit $|\varphi|={{|U_N|}\over{c^2}}\ll
1$. However, additional differences may also arise in the strong
field limit as with the exponential metric $\lambda=e^{|\varphi|}$,
$ {\cal{N}}= e^{2|\varphi|}$, see \cite{ultraweak}.}. Our scope here
is to refine and update the analysis of \cite{gerg} namely : i) in
principle, one expects a non-zero fractional anisotropy  ${\cal
O}(10^{-15})$ of the velocity of light in the vacuum ii) if the
vacuum is considered a quantum medium there are non-trivial
implications for the analysis of the experiments iii) most probably,
this tiny effect is now observed in many room-temperature
ether-drift experiments and, in particular, in a very precise
cryogenic experiment \cite{cpt2013} where the level ${\cal
O}(10^{-15})$ is now about 100 times larger than the designed
short-term stability.

\vskip 15pt

{\bf 2.}~ For the problem of measuring the speed of light, we shall
follow closely ref.\cite{gerg} (to which we address the reader for
more details). The main point is that, to determine speed as
(distance moved)/(time taken), one must first choose some standards
of distance and time. Since different choices can give different
answers, we shall adopt the point of view of special relativity: the
right space-time units are those for which the speed of light in the
vacuum $c_\gamma$, when measured in an inertial frame, coincides
with the basic parameter $c$ entering Lorentz transformations.
However, inertial frames are just an idealization. Therefore the
appropriate realization is to assume {\it local} standards of
distance and time such that the identification $c_\gamma=c$ holds as
an asymptotic relation in the physical conditions which are as close
as possible to an inertial frame, i.e. {\it in a freely falling
frame} (at least by restricting to a space-time region small enough
that tidal effects of the external gravitational potential $U_{\rm
ext}(x)$ can be ignored). This is essential to obtain an operative
definition of the otherwise unknown parameter $c$.

With these premises, light propagation for an observer $S'$ sitting
on the Earth's surface can be described with increasing degrees of
approximations:

\begin{figure}
\begin{center}
\epsfig{figure=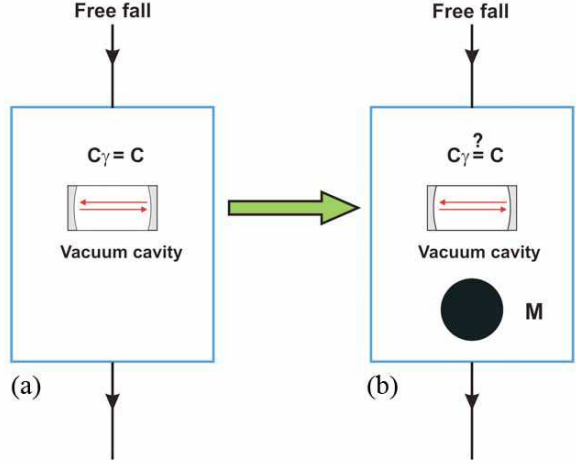,width=8.5truecm,angle=0}
\end{center}
\caption{ {\it A pictorial representation of the effect of a heavy
mass $M$ carried on board of a freely-falling system, case (b). With
respect to case (a), in a flat-space picture of gravity, the mass
$M$ modifies the effective, local space-time structure by re-scaling
the physical units ($dx$, $dy$, $dz$, $dt$) $\to$ ($d\hat{x}$,
$d\hat{y}$, $d\hat{z}$, $d\hat{t}$) and introducing a non-trivial
refractive index ${\cal N}\neq 1$ so that now $c_\gamma \neq c$. The
figure is taken from ref.\cite{chaos}.}
 } \label{Fig.1}
\end{figure}

~~~i) $S'$ is considered a freely falling frame. This amounts to
assume $c_\gamma=c$ so that, given two events which, in terms of the
local space-time units of $S'$, differ by $(dx, dy, dz, dt)$, light
propagation is described by the condition (ff='free-fall')
\BE\label{zero1} (ds^2)_{\rm ff}=c^2dt^2- (dx^2+dy^2+dz^2)=0~\EE
~~~ii) To a closer look, however, an observer  $S'$ placed on the
Earth's surface can only be considered a freely-falling frame up to
the presence of the Earth's gravitational field. Its inclusion  can
be estimated by considering $S'$ as a freely-falling frame, in the
same external gravitational field described by $U_{\rm ext}(x)$,
that however is also carrying on board a heavy object of mass $M$
(the Earth's mass itself) that affects the effective local
space-time structure, see Fig.1. To derive the required correction,
let us again denote by ($dx$, $dy$, $dz$, $dt$) the local space-time
units of the freely-falling observer $S'$ in the limit $M=0$ and by
$\delta U$ the extra Newtonian potential produced by the heavy mass
$M$ at the experimental set up where one wants to describe light
propagation. From Eqs.(\ref{first}) and (\ref{second}), light
propagation for the $S'$ observer is now described by re-scaled
units ($d\hat{x}$, $d\hat{y}$, $d\hat{z}$, $d\hat{t}$) and a
refractive index as \BE\label{iso}(ds^2)_{\rm \delta U}
={{c^2d\hat{t} ^2}\over{{\cal N}^2 }}-
(d\hat{x}^2+d\hat{y}^2+d\hat{z}^2)=0~\EE where, to first order in
$\delta U$, the re-scaling $\lambda$ and ${\cal N}$ are  \BE
\label{lambda} \lambda \sim 1+{{|\delta U|}\over{c^2}}~~~~~~
~~~~~~~~~~~~ ~~~~~~ {\cal N}\sim 1+2{{|\delta U|}\over{c^2}} \EE As
anticipated, to this order, the metric is formally as in general
relativity \BE\label{gr} (ds^2)_{\rm GR}=c^2dT^2(1-2{{|U_{\rm
N}|}\over{c^2}})- (dX^2+dY^2+dZ^2)(1+2{{|U_{\rm
N}|}\over{c^2}})\equiv c^2 d\tau^2 - dl^2\EE where $U_N$ denotes the
Newtonian potential and ($dT$, $dX$, $dY$, $dZ$) arbitrary
coordinates defined for $U_{\rm N}=0$. Finally, $d\tau$ and $dl$
denote the elements of proper time and proper length in terms of
which, in general relativity, one would again deduce from $ds^2=0$
the same universal value $c={{dl}\over{d\tau}}$. This is the basic
difference with Eqs.(\ref{iso}), (\ref{lambda}) where the physical
unit of length is $\sqrt {d\hat{x}^2+d\hat{y}^2+d\hat{z}^2}$, the
physical unit of time is $d\hat{t}$ and  instead a non-trivial
refractive index ${\cal N}$ is introduced. For an observer placed on
the Earth's surface, its value is \BE \label{refractive}{\cal N}- 1
\sim {{2G_N M}\over{c^2R}} \sim 1.4\cdot 10^{-9}\EE  $M$ and $R$
being respectively the Earth's mass and radius.

~~~iii) Differently from general relativity, in a flat-space
interpretation with re-scaled units ($d\hat{x}$, $d\hat{y}$,
$d\hat{z}$, $d\hat{t}$) and ${\cal N}\neq 1$, the speed of light in
the vacuum $c_\gamma$ no longer coincides with the parameter $c$
entering Lorentz transformations. Therefore, as a general
consequence of Lorentz transformations, an isotropic propagation as
in Eq.(\ref{iso}) can only be valid for a special state of motion of
the Earth's laboratory. This provides the {\it operative definition
of a preferred reference frame} $\Sigma$ while for a non-zero
relative velocity ${\bf V}$  there are off diagonal elements
$g_{0i}\neq 0$ in the effective metric and a tiny light anisotropy.
These off diagonal elements can be imagined as being due to a
directional polarization of the vacuum induced by the now moving
Earth's gravitational field and express the general property
\cite{volkov} that any metric, locally, can always be brought into
diagonal form by suitable rotations and boosts. As shown in
ref.\cite{gerg}, to first order in both $({\cal N}- 1)$ and $V/c$
one finds \BE g_{0i}\sim 2({\cal N}- 1){{V_i}\over{c}} \EE In this
way, by introducing $\beta=V/c$, $\kappa=( {\cal N}^2 -1)$ and the
angle $\theta$ between ${\bf V}$ and the direction of light
propagation, one finds, to ${\cal O}(\kappa)$ and ${\cal
O}(\beta^2)$, the one-way velocity \cite{gerg}
\BE \label{oneway}
       c_\gamma(\theta)= {{c} \over{{\cal N}}}~\left[
       1- \kappa \beta \cos\theta -
       {{\kappa}\over{2}} \beta^2(1+\cos^2\theta)\right]
\EE and a two-way velocity
\begin{eqnarray}
\label{twoway}
       \bar{c}_\gamma(\theta)&=&
       {{ 2  c_\gamma(\theta) c_\gamma(\pi + \theta) }\over{
       c_\gamma(\theta) + c_\gamma(\pi + \theta) }} \nonumber \\
       &\sim& {{c} \over{{\cal N}}}~\left[1-\kappa\beta^2\left(1 -
       {{1}\over{2}} \sin^2\theta\right) \right]
\end{eqnarray}
This gives finally \footnote{There is a subtle difference between
our Eqs.(\ref{oneway}) and(\ref{twoway}) and the corresponding Eqs.
(6) and (10) of Ref.~\cite{pla} that has to do with the relativistic
aberration of the angles. Namely, in Ref.\cite{pla}, with the
(wrong) motivation that the anisotropy is ${\cal O}(\beta^2)$, no
attention was paid to the precise definition of the angle between
the Earth's velocity and the direction of the photon momentum. Thus
the two-way speed of light in the $S'$ frame was parameterized in
terms of the angle $\theta\equiv\theta_\Sigma$ as seen in the
$\Sigma$ frame. This can be explicitly checked by replacing in our
Eqs.~(\ref{oneway}) and(\ref{twoway}) the aberration relation
$\cos \theta_{\rm lab}=(-\beta + \cos\theta_\Sigma)/
       (1-\beta\cos\theta_\Sigma)$
or equivalently by replacing $\cos \theta_{\Sigma}=(\beta +
\cos\theta_{\rm lab})/ (1+\beta\cos\theta_{\rm lab})$ in Eqs. (6)
and (10) of Ref.~\cite{pla}. However, the apparatus is at rest in
the laboratory frame, so that the correct orthogonality condition of
two optical cavities at angles $\theta$ and $\pi/2 + \theta$ is
expressed in terms of $\theta=\theta_{\rm lab}$ and not in terms of
$\theta=\theta_{\Sigma}$. This trivial remark produces however a
non-trivial difference. In fact, the final anisotropy Eq.
(\ref{rms}) is now smaller by a factor of 3 than the one computed in
Ref.\cite{pla} by adopting the wrong definition of orthogonality in
terms of $\theta=\theta_{\Sigma}$.} \BE \label{rms}
       {{\bar{c}_\gamma(\pi/2 +\theta)- \bar{c}_\gamma (\theta)} \over
       {\langle \bar{c}_\gamma \rangle }} \sim ({\cal N}-1)
       {{v^2 }\over{c^2}} \cos2(\theta-\theta_0) \EE
where the pair $(v,\theta_0)$ describes the projection of ${\bf{V}}$
onto the relevant plane. From the previous analysis, by using
Eq.(\ref{refractive}) and adopting, as a rough order of magnitude,
the typical value of most cosmic motions $V\sim$ 300 km/s, one thus
expects a tiny fractional anisotropy \BE \label{averani}
        {{\langle\Delta \bar{c}_\theta \rangle} \over{c}} \sim
       ({\cal N}-1){{V^2 }\over{c^2}} ={\cal O}(10^{-15}) \EE
that could finally be detected in the present, precise ether-drift
experiments.

\vskip 15pt

{\bf 3.}~In present ether-drift experiments one measures the
frequency shift, i.e. the beat signal, $\Delta \nu$ of two rotating
optical resonators whose definite non-zero value would provide a
direct measure of an anisotropy of the velocity of light
\cite{applied}. In this framework, the possible time modulation of
the signal that might be induced by the Earth's rotation (and its
orbital revolution) has always represented a crucial ingredient for
the analysis of the data. To see this, let us re-write
Eq.(\ref{rms}) as \BE \label{basic2}
    {{\Delta \nu (t)}\over{\nu_0}} = {{\Delta \bar{c}_\theta(t) } \over{c}}
    \sim
 ({\cal N}-1){{v^2(t) }\over{c^2}}\cos 2(\omega_{\rm rot}t
-\theta_0(t)) \EE where $\nu_0$ indicates the reference frequency of
the two resonators and $\omega_{\rm rot}$ is the rotation frequency
of the apparatus. Therefore one finds \BE \label{basic3} {{\Delta
\nu (t)}\over{\nu_0}} \sim 2{S}(t)\sin 2\omega_{\rm rot}t +
      2{C}(t)\cos 2\omega_{\rm rot}t\EE with \BE \label{amplitude10}
       C(t)= {{1}\over{2}}({\cal N}-1)~ {{v^2_x(t)- v^2_y(t)  }
       \over{c^2}}~~~~~~~~~~~~~~~~~~~~ S(t)={{1}\over{2}}({\cal N}-1) ~{{2v_x(t)v_y(t)  }\over{c^2}}
\EE and $v_x(t)=v(t)\cos\theta_0(t)$, $v_y(t)=v(t)\sin\theta_0(t)$

The standard assumption to analyze the data is to consider a cosmic
Earth's velocity with well defined magnitude $V$, right ascension
$\alpha$ and angular declination $\gamma$ that can be considered
constant for short-time observations of a few days where there are
no appreciable changes due to the Earth's orbital velocity around
the Sun. In this framework, where the only time dependence is due to
the Earth's rotation, one identifies $v(t)\equiv \tilde v(t)$ and
$\theta_0(t)\equiv\tilde\theta_0(t)$ where $\tilde v(t)$ and
$\tilde\theta_0(t)$ derive from the simple application of spherical
trigonometry \cite{gerg} \BE \label{nassau1}
       \cos z(t)= \sin\gamma\sin \phi + \cos\gamma
       \cos\phi \cos(\tau-\alpha)
\EE \BE \label{projection}
       \tilde {v}(t) =V \sin z(t)
\EE \BE\label{nassau2}
      \tilde{v}(t)\cos\tilde\theta_0(t)= V\left[ \sin\gamma\cos \phi -\cos\gamma
       \sin\phi \cos(\tau-\alpha)\right]
\EE \BE\label{nassau3}
      \tilde{v}(t)\sin\tilde\theta_0(t)= V\cos\gamma\sin(\tau-\alpha) \EE
  Here $z=z(t)$ is the zenithal distance of ${\bf{V}}$, $\phi$ is
the latitude of the laboratory, $\tau=\omega_{\rm sid}t$ is the
sidereal time of the observation in degrees ($\omega_{\rm sid}\sim
{{2\pi}\over{23^{h}56'}}$) and the angle $\theta_0$ is counted
conventionally from North through East so that North is $\theta_0=0$
and East is $\theta_0=90^o$. In this way,  one finds
\BE\label{amorse1}
      S(t)\equiv {\tilde S}(t) =
      {S}_{s1}\sin \tau +{S}_{c1} \cos \tau
       + {S}_{s2}\sin(2\tau) +{S}_{c2} \cos(2\tau)
\EE \BE \label{amorse2}
      C(t)\equiv {\tilde C}(t) = {C}_0 +
      {C}_{s1}\sin \tau +{C}_{c1} \cos \tau
       + {C}_{s2}\sin(2 \tau) +{C}_{c2} \cos(2 \tau)
\EE In this picture, the  $C_k$ and $S_k$ Fourier coefficients
depend on the three parameters $(V,\alpha,\gamma)$ (see
\cite{applied}) and, to very good approximation, should be
time-independent for short-time observations. Thus, by accepting
this theoretical framework, it becomes natural to average the
various $C_k$ and $S_k$ obtained from fits performed during a 1$-$2
day observation period. In this case, although the typical
instantaneous $S(t)$ and $C(t)$ are ${\cal O}(10^{-15})$, due to the
irregular nature of the observed signal, there are strong
cancelations with global averages $(C_k)^{\rm avg}$ and $(S_k)^{\rm
avg}$ for the Fourier coefficients at the level ${\cal O}(10^{-17})$
\cite{newberlin,newschiller}. This is usually taken as an indication
that the much larger instantaneous signal is a spurious instrumental
effect, e.g. thermal noise.

However, there might be forms of ether-drift where the
straightforward parameterizations Eqs.(\ref{amorse1}),
(\ref{amorse2}) and the associated averaging procedures are {\it
not} allowed. For this reason, before assuming any definite
theoretical scenario, one could first ask: if light were really
propagating in a physical medium, an ether, and not in a trivial
empty vacuum, how should the motion of (or in) this medium be
described? Namely, could this relative motion exhibit variations
that are {\it not} only due to known effects as the Earth's rotation
and orbital revolution? Here, there is the logical gap. In fact, by
comparing the Earth's cosmic motion with that of a body in a fluid,
the standard picture Eqs.(\ref{nassau1})$-$(\ref{amorse2}) amounts
to the condition of a pure laminar flow where global and local
velocity fields coincide. Instead, the relation between the
macroscopic Earth's motion and the measurements performed in a
laboratory depends on the physical nature of the vacuum. If we
consider the vacuum as a superfluid, i.e. a quantum liquid, then the
frequency shifts will likely exhibit the typical irregular,
non-deterministic pattern which characterizes any quantum
measurement. In this case, in view of the striking similarities
\cite{vinen} between many aspects of turbulence in fluids and
superfluids, and of the intriguing derivation of Kolmogorov scaling
laws \cite{kolmo} from quantum hydrodynamics \cite{yogi}, rather
than adopting the simple classical model of a laminar flow, one
could try to compare the experimental data with models of a {\it
turbulent} flow, see Fig.2. In this alternative scenario, the same
basic experimental data might admit a different interpretation and a
genuine stochastic signal $\Delta \nu (t)\neq 0$ could perfectly
coexist with $(C_k)^{\rm avg} \sim (S_k)^{\rm avg}\sim 0$.

\begin{figure}[ht]
\begin{center}
\epsfig{figure=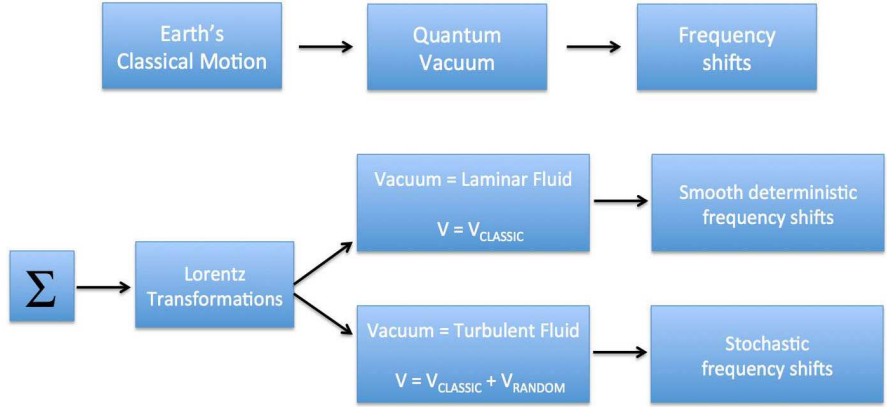,height=6 true cm,width=12 true
cm,angle=0}
\end{center}
\caption{\it The two possible ways to relate Earth's classical
motion and frequency shifts. }
\end{figure}

\vskip 15pt

{\bf 4.}~By considering the vacuum as a fluid, one is naturally
driven to the limit of zero viscosity where the local velocity field
becomes non-differentiable and the ordinary formulation in terms of
differential equations becomes inadequate \cite{onsager}. Thus, one
has to adopt some other description, for instance a formulation in
terms of random Fourier series \cite{onsager,landau}. In this other
approach, the parameters of the macroscopic motion are only used to
fix the limiting boundaries \cite{fung} for a microscopic velocity
field which has instead an intrinsic stochastic nature.
\begin{figure}[ht]
\begin{center}
\epsfig{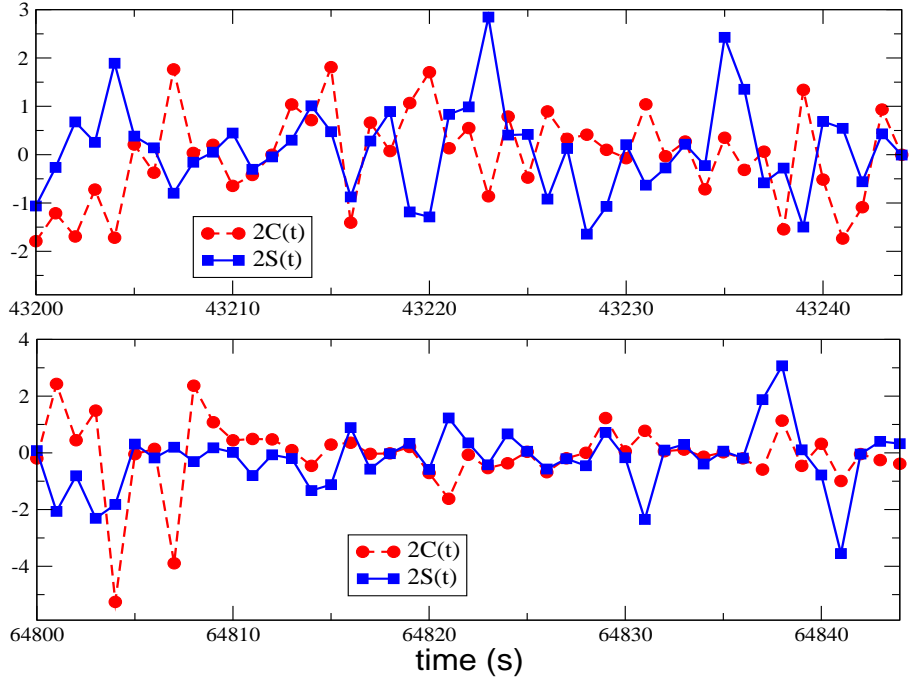}
\end{center}
\caption{\it Two typical sets of 45 seconds for the instantaneous
$2C(t)$ and $2S(t)$  in units $10^{-15}$. The two sets belong to the
same random sequence and refer to two sidereal times that differ by
6 hours. The boundaries of the stochastic velocity components in
Eqs.(\ref{vx}) and (\ref{vy}) are controlled by
$(V,\alpha,\gamma)_{\rm CMB}$ through Eqs.(\ref{projection}) and
(\ref{isot}).}
\end{figure}
The simplest choice, adopted in ref.\cite{physica}, corresponds to a
turbulence which, at small scales, appears statistically homogeneous
and isotropic \footnote{This picture reflects the basic Kolmogorov
theory \cite{kolmo} of a fluid with vanishingly small viscosity.}.
This represents a zeroth-order approximation but, nevertheless, it
is  useful to illustrate basic phenomenological features associated
with an underlying stochastic vacuum. The perspective is that of an
observer moving in the turbulent fluid who wants to simulate the two
components of the velocity in his x-y plane, at a given fixed
location in his laboratory, to reproduce the $S(t)$ and $C(t)$
functions Eq.(\ref{amplitude10}). In a statistically homogeneous
turbulence, one finds the general expressions \BE \label{vx} v_x(t)=
\sum^{\infty}_{n=1}\left[
       x_n(1)\cos \omega_n t + x_n(2)\sin \omega_n t \right] \EE
\BE \label{vy} v_y(t)= \sum^{\infty}_{n=1}\left[
       y_n(1)\cos \omega_n t + y_n(2)\sin \omega_n t \right] \EE
where $\omega_n=2n\pi/T$, T being a time scale which represents a
common period of all stochastic components. For numerical
simulations, the typical value $T=T_{\rm day}$= 24 hours was adopted
\cite{physica}. However, it was also checked with a few runs that
the statistical distributions of the various quantities do not
change substantially by varying $T$ in the rather wide range
$0.1~T_{\rm day}\leq T \leq 10~T_{\rm day}$.

The coefficients $x_n(i=1,2)$ and $y_n(i=1,2)$ are random variables
with zero mean and have the physical dimension of a velocity.
Without necessarily assuming statistical isotropy, let us denote by
$[-\tilde {v}_x(t),\tilde {v}_x(t)]$ the range for $x_n(i=1,2)$ and
by $[-\tilde {v}_y(t),\tilde {v}_y(t)]$ the corresponding range for
$y_n(i=1,2)$. In terms of these boundaries, the only non-vanishing
(quadratic) statistical averages are \BE \label{quadratic} \langle
x^2_n(i=1,2)\rangle_{\rm stat}={{{\tilde v}^2_x(t) }\over{3
~n^{2\eta}}}~~~~~~~~~~~~~~~~~~~~~~~~~~~~\langle
y^2_n(i=1,2)\rangle_{\rm stat}={{{\tilde v}^2_y(t) }\over{3
~n^{2\eta}}} \EE  in a uniform probability model within the
intervals $[-\tilde {v}_x(t),\tilde {v}_x(t)]$ and $[-\tilde
{v}_y(t),\tilde {v}_y(t)]$. Here, the exponent $\eta$ controls the
power spectrum of the fluctuating components. For numerical
simulations, between the two values $\eta=5/6$ and $\eta=1$ reported
in ref.\cite{fung}, we have adopted $\eta=1$ which corresponds to
the point of view of an observer moving in the fluid.

As definite boundaries one could choose for instance $\tilde
v_x(t)\equiv \tilde v(t) \cos \tilde\theta_0(t)$, $\tilde
v_y(t)\equiv \tilde v(t) \sin \tilde\theta_0(t)$, $\tilde v(t)$ and
$\tilde\theta_0(t)$ being defined in Eqs.
(\ref{nassau1})$-$(\ref{projection}). In this case, the set $V=$ 370
km/s, $\alpha=168$ degrees, $\gamma= -6$ degrees, which describes
the average Earth's motion with respect to the CMB, was shown
\cite{plus} to provide a good statistical description of Joos'1930
observations \cite{joos} whose fringe-shift amplitudes $A(t)$,
differently from the phases $\theta_0(t)$, can be extracted
unambiguously from the original article. Finally, while still
preserving $\tilde{v}^2_x(t) + \tilde{v}^2_y(t)=\tilde{v}^2(t)$, one
could enforce statistical isotropy, in agreement with Kolmogorov's
theory, by choosing the common value from Eq.(\ref{projection}) \BE
\label{isot} \tilde{v}_x(t)=\tilde{v}_y(t)={{
\tilde{v}(t)}\over{\sqrt{2} }} \EE For such isotropic model, by
combining Eqs.(\ref{vx})$-$(\ref{isot}) one gets \BE
\label{vanishing} \langle v^2_x(t)\rangle_{\rm stat}=\langle
v^2_y(t)\rangle_{\rm stat}={{\tilde{v}^2(t)}\over{2}}~{{1}\over{3}}
\sum^{\infty}_{n=1} {{1}\over{n^2}}~~~~~~~~~~~~~~~~~~~~~~~~ \langle
v_x(t)v_y(t)\rangle_{\rm stat}=0 \EE with vanishing statistical
averages $\langle C(t)\rangle_ {\rm stat}=0$ and $\langle
S(t)\rangle_ {\rm stat}=0$ at any time $t$.

\begin{figure}
\begin{center}
\psfig{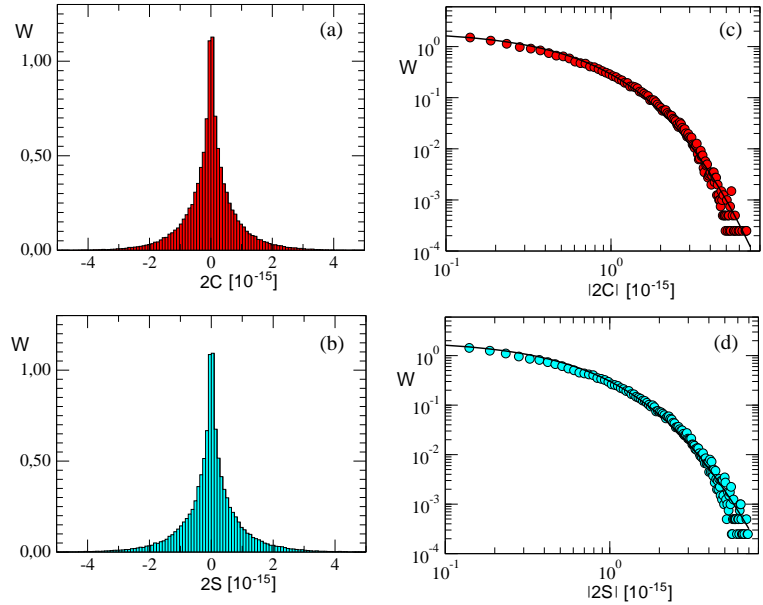}
\end{center}
\caption{ {\it We show, see (a) and (b), the histograms $W$ obtained
from a single simulation of measurements of $\rm 2C=2C(t)$ and $\rm
2S=2S(t)$ performed at regular steps of 1 second over an entire
sidereal day . The vertical normalization is to a unit area. The
mean values are $\rm \langle 2C\rangle_{\rm day} =-1.6 \cdot
10^{-18}$, $\rm \langle 2S\rangle_{\rm day} =4.3 \cdot 10^{-18}$ and
the standard deviations $\rm \sigma_{\rm day}(2C)=0.87 \cdot
10^{-15}$, $\rm \sigma_{\rm day}(2S)=0.96\cdot 10^{-15}$. We also
show, see (c) and (d), the corresponding plots in a log-log scale
and the fits with Eq.(\ref{qexp}). The boundaries of the stochastic
velocity components in Eqs.(\ref{vx}) and (\ref{vy}) are controlled
by $(V,\alpha,\gamma)_{\rm CMB}$ through Eqs.(\ref{projection}) and
(\ref{isot}).}} \label{Fig.4}
\end{figure}

To have an idea of the signal in this case, we report in Fig.3 two
typical sets of the instantaneous values for $2C(t)$ and $2S(t)$
during one rotation period $T_{\rm rot}=$ 45 seconds of the
apparatus of ref.\cite{newberlin}. The two sets belong to the same
random sequence and refer to two sidereal times that differ by 6
hours. As in \cite{plus}, the set $(V,\alpha,\gamma)_{\rm CMB}$ was
adopted to control the boundaries of the stochastic velocity
components through Eqs.(\ref{projection})and (\ref{isot}). The value
$\phi= 52$ degrees was also fixed to reproduce the average latitude
of the laboratories in Berlin and D\"usseldorf.

We have also simulated long sequences of measurements performed at
regular steps of 1 second over an entire sidereal day. For a
particular random sequence, the resulting histograms of $2C$ and
$2S$ are reported in panels (a) and (b) of Fig.4. Notice that these
distributions are clearly ``fat-tailed'' and very different from a
Gaussian shape. This kind of behavior is characteristic of
probability distributions for instantaneous data in turbulent flows
(see e.g. \cite{sreenivasan,beck}). To better appreciate the
deviation from Gaussian behavior, in panels (c) and (d) we plot the
same data in a log$-$log scale. The resulting distributions are well
fitted by the so-called $q-$exponential function \cite{tsallis} \BE
\label{qexp} f_q(x) = a (1 - (1 - q) x b)^{1 / (1 - q)}\EE  with
entropic index $q\sim 1.1$. In view of Eqs.(\ref{vanishing}) any
non-zero average $ \langle 2C\rangle_{\rm day}$ and $ \langle
2S\rangle_{\rm day}$ should be considered as statistical
fluctuation.  On the other hand, the standard deviations
$\sigma(2C)$ and $\sigma(2S)$ have definite non-zero values \BE
\label{sigmas} \sigma_{\rm day}(2C)\sim (0.87 \pm 0.08)\cdot
10^{-15}
 ~~~~~~~~~~~~~~~~~~ \sigma_{\rm day}(2S)\sim(0.96 \pm 0.09)\cdot 10^{-15}\EE
where uncertainties reflect the observed variations due to the
truncation of the Fourier modes in Eqs.(\ref{vx}), (\ref{vy}) and to
the dependence on the random sequence.

Another reliable indicator is the statistical average of the
amplitude of the signal $A(t)\equiv 2\sqrt{S^2(t) + C^2(t)} $. In
this case, by using Eqs. (\ref{refractive}) and (\ref{vanishing}),
one finds \footnote{Notice that, as far as the amplitude of the
signal is concerned, the isotropic model Eq.(\ref{isot}) cannot be
distinguished from the non-isotropic choice $\tilde v_x(t)\equiv
\tilde v(t) \cos \tilde\theta_0(t)$ and $\tilde v_y(t)\equiv \tilde
v(t) \sin \tilde\theta_0(t)$. For this reason, the statistical
analysis of Joos' amplitude data \cite{plus} would remain the same
in the two models. } \BE \label{reduction}\langle A(t) \rangle_{\rm
stat}=({\cal N}-1)
 {{\tilde{v}^2(t)}\over{c^2}}
~{{1}\over{3}} \sum^{\infty}_{n=1} {{1}\over{n^2}}=
{{\pi^2}\over{18}}~1.4\cdot 10^{-15}
{{\tilde{v}^2(t)}\over{(300~{\rm km/s})^2}} \EE By maintaining the
CMB parameters $(V,\alpha,\gamma)_{\rm CMB}$ and fixing $\phi= 52$
degrees, one gets a daily average $\sqrt {\langle \tilde {v}^2
\rangle}_{\rm day} \sim 332$ km/s from the relation \cite{gerg} \BE
\langle \tilde{v}^2 \rangle_{\rm day}= V^2
       \left(1- \sin^2\gamma\sin^2\phi
%     \right.
%      \nonumber \\  \left.\right.
%      & & - \left.
       - {{1}\over{2}} \cos^2\gamma\cos^2\phi \right) \EE
In this way, one predicts an average amplitude $\langle A
\rangle_{\rm day}\sim 10^{-15}$. Notice however that, by performing
extensive simulations, there are occasionally large spikes of the
instantaneous amplitude, up to $(6\div 7) \cdot 10^{-15}$, when many
Fourier modes sum up coherently (see the tails in panels (c) and (d)
of Fig.4). The effect of these spikes gets smoothed when averaging
but their presence is characteristic of a stochastic model. With a
standard attitude, where one only expects smooth time modulations,
the observation of such spikes would naturally be interpreted as a
spurious disturbance. More precise tests of the model could be
performed if real data for $A(t)$, $S(t)$ and $C(t)$ will become
available.

\vskip 15pt

{\bf 5.}~An instantaneous stochastic signal $\sim 10^{-15}$ is well
consistent with the most precise room temperature experiments
\cite{newberlin,newschiller}. Since this observed value is
comparable to the theoretical estimate of ref.\cite{numata}, this
has been interpreted in terms of thermal noise in the mirrors and
the spacers of the optical resonators. However, as pointed out in
ref.\cite{gerg}, this interpretation is not unique because roughly
the same value is also obtained from {\it cryogenic} experiments
\cite{muller,antonini,cpt2013}. The point is that the standard
estimate of thermal disturbances \cite{numata} is based on the
fluctuation-dissipation theorem, and therefore there is no obvious
reason why room temperature and cryogenic resonators should exhibit
the same instrumental effects. The unexplained agreement with the
very recent result of ref.\cite{cpt2013} is particularly striking in
view of the factor 100 which now exists between observed stochastic
signal $10^{-15}$ and designed short-term stability ${\cal O}(
10^{-17})$. Tentatively, the authors of \cite{cpt2013} interpret
this discrepancy in terms of a lack of rigidity of their cryostat.
However, probably, they have not considered the possibility of a
genuine random signal and of intrinsic limitations placed by the
vacuum structure. In this different perspective, the interpretation
proposed here should also be taken into account for its ultimate
implications.

\end{document}